\documentclass{PoS}
\usepackage{amsmath}

\title{$B$ decays to radially excited $D$}

\ShortTitle{}

\author{Damir Becirevic, \speaker{Beno\^\i t Blossier}, Antoine G\'erardin, Alain Le Yaouanc, 
	   Francesco Sanfilippo\\
        Laboratoire de Physique Th\'eorique, CNRS/Universit\'e Paris-Sud XI, F-91405 Orsay Cedex, France\\
        E-mail: \email{benoit.blossier@th.u-psud.fr}}

\abstract{We discuss the possibility to measure in present experiments, especially LHCb, the non leptonic decay branching ratio $B \to D' \pi$, and emphasize phenomenological implications on $B \to D' l \nu$ semileptonic decay. We have estimated by lattice QCD the $D'$ decay constant $f_{D'}$ that parameterizes
the $D'$ emission contribution to the Class-III non leptonic decay $B^- \to D^0 \pi^-$. In addition,
we provide a new estimate of the decay constants $f_{D_{s,q}}$ which read $f_{D_{s}}=252(3)$~MeV and $f_{D_{s}}/f_{D}=1.23(1)(1)$.}

\FullConference{31st International Symposium on Lattice Field Theory - LATTICE 2013\\
		July 29 - August 3, 2013\\
		Mainz, Germany}

\begin{document}

\section{Introduction}

\begin{figure}
\begin{center}
\begin{tabular}{cc}
\begin{tabular}{c}
\includegraphics*[width=6cm, height=3.75cm]{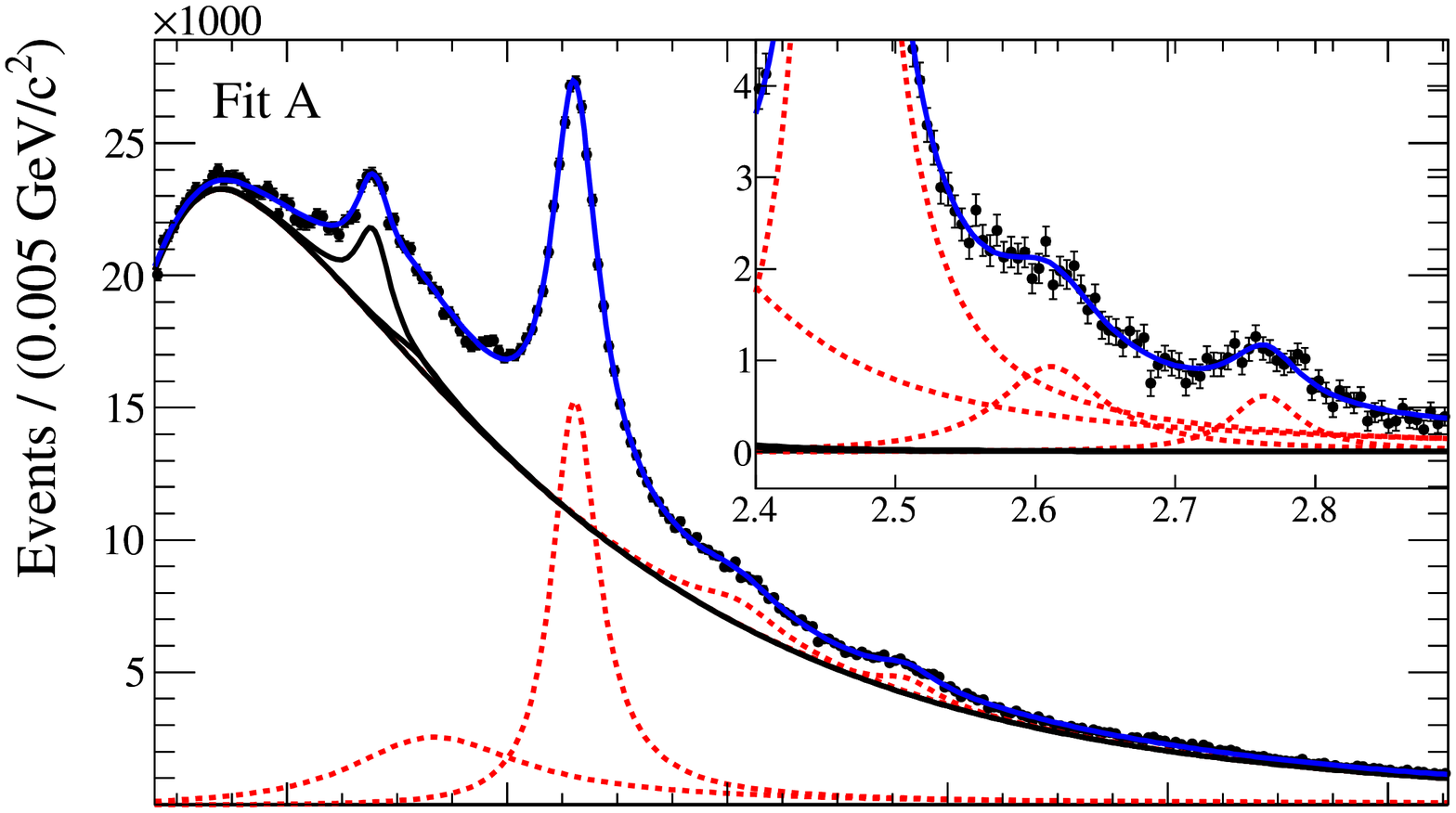}\\
\includegraphics*[width=6cm, height=3.75cm]{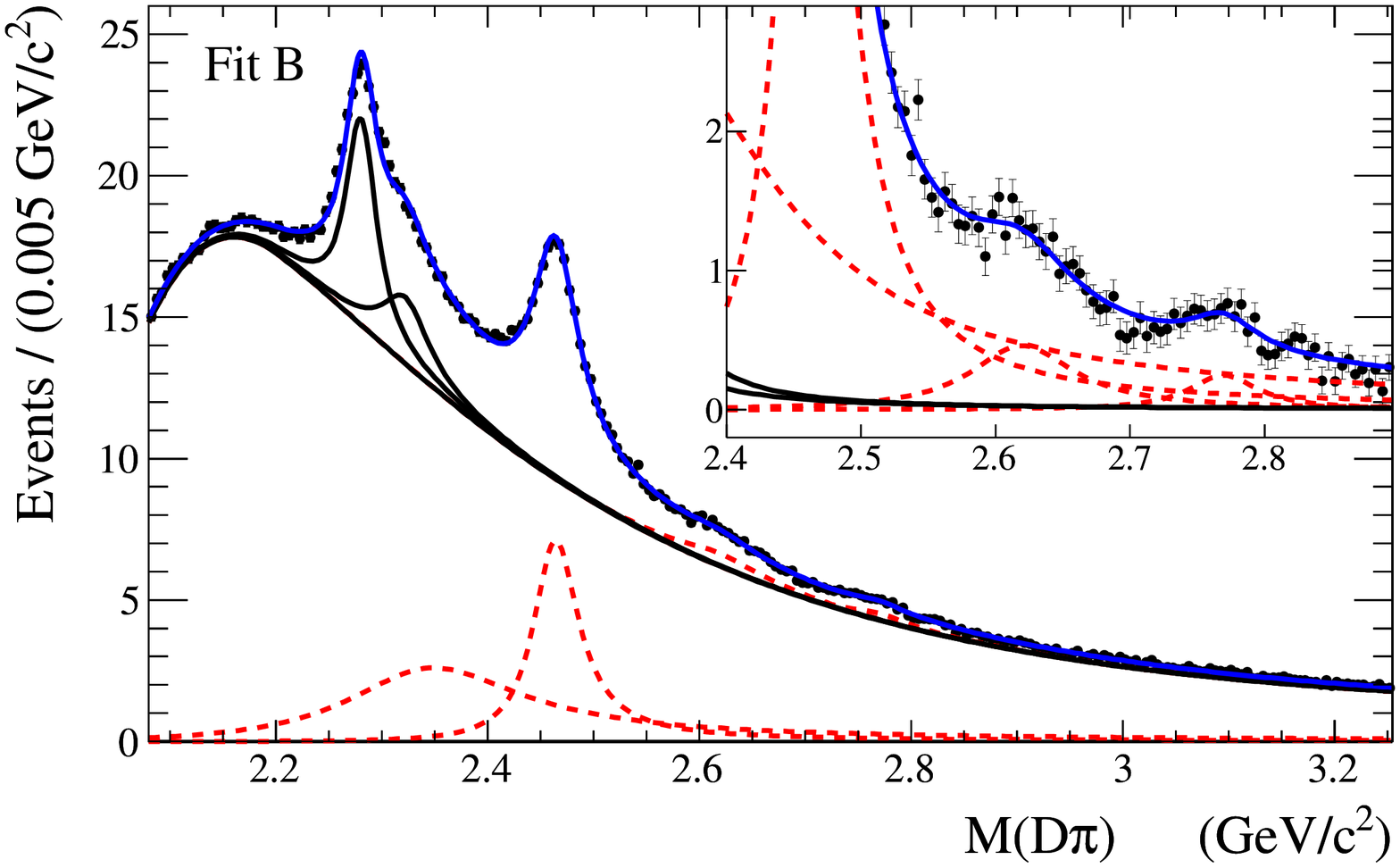}
\end{tabular}
&
\begin{tabular}{c}
\includegraphics*[width=6cm, height=2.5cm]{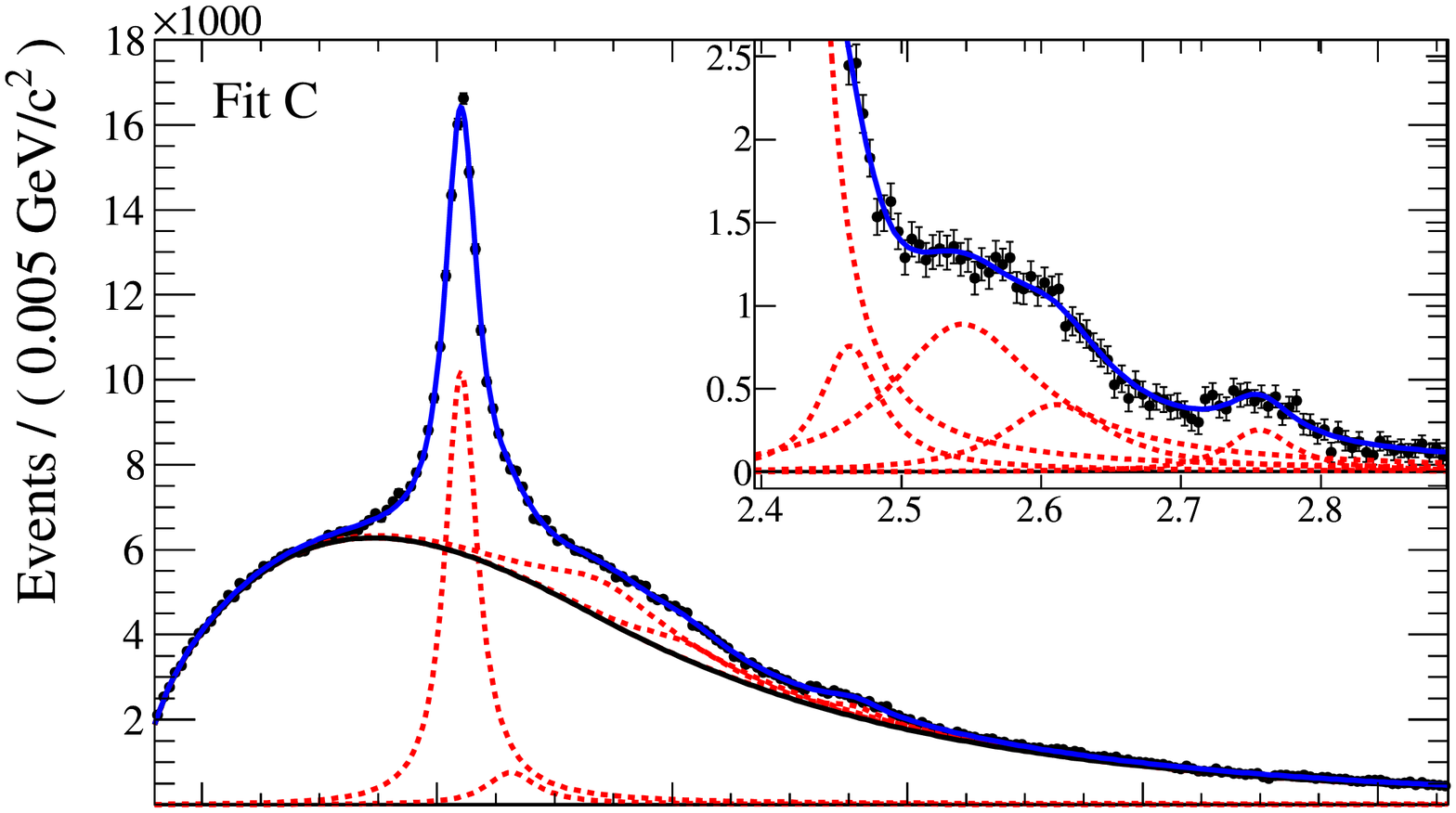}\\
\includegraphics*[width=6cm, height=2.5cm]{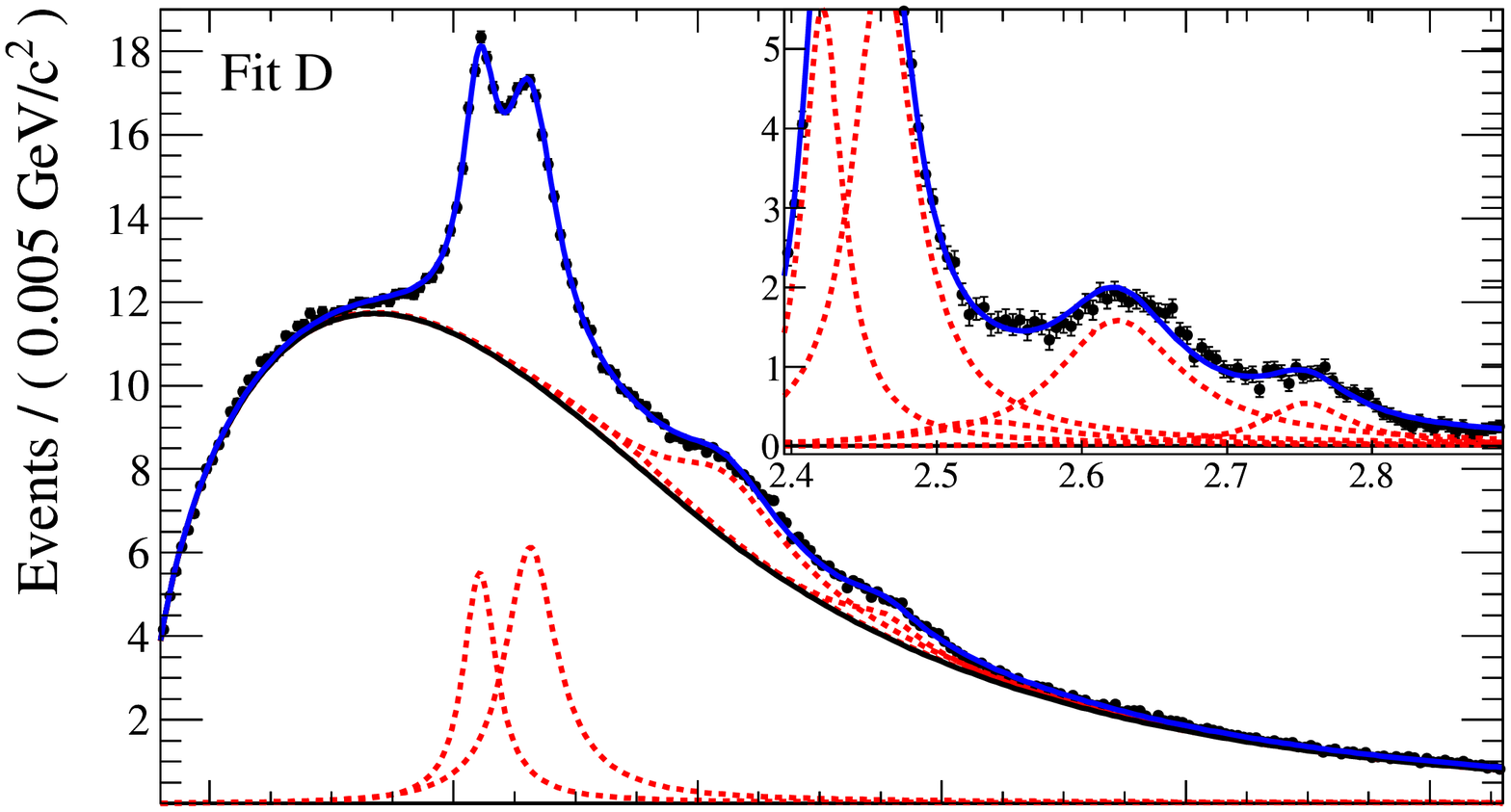}\\
\includegraphics*[width=6cm, height=2.5cm]{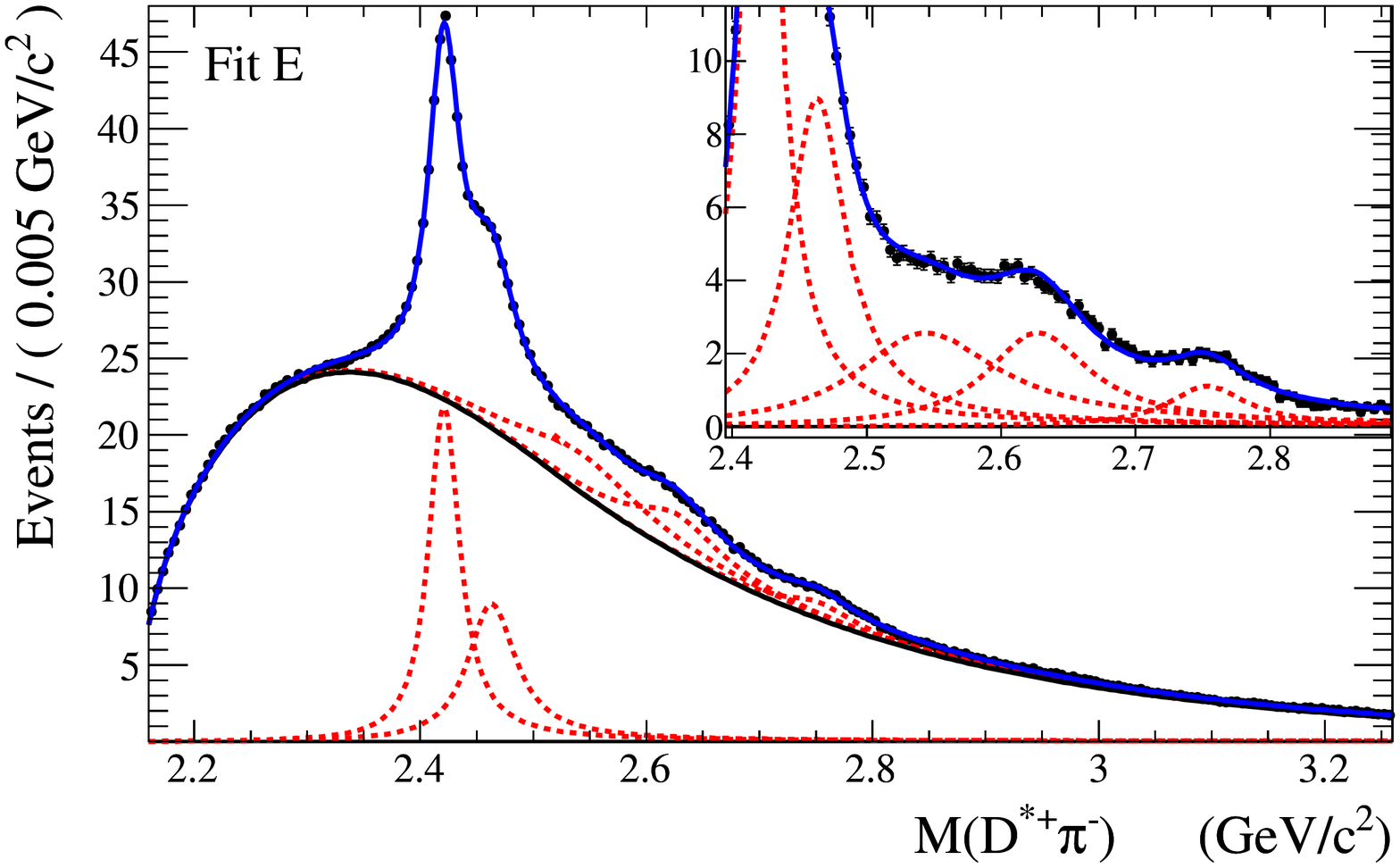}
\end{tabular}
\end{tabular}
\end{center}
\caption{\label{figDspectrum} $D\pi$ (left panel) and $D^*\pi$ (right panel) mass distributions obtained at Babar \cite{babard}.}
\end{figure}

Understanding the long-distance dynamics of QCD is of key importance to control the theoretical systematics on low-energy processes that are investigated at LHC in order to detect indirect effects of New Physics. With that respect beauty and charmed mesons represents a particularly rich sector. Recently, the Babar Collaboration claimed to have isolated a bench of new $D$ states \cite{babard}. $D \pi$ and $D^* \pi$ mass distributions are depicted in Figure \ref{figDspectrum}: in the former, one observes a clear peak, corresponding to the state $D^*_2(2460)$, and two "enhancements" that are interpreted as states $D^*(2600)$ and $D^*(2760)$. In the latter, one observes a peak at $D_1(2420)$ and distinguishes two structures that are interpreted as $D(2550)\equiv D'$ and $D(2750)$.
Performing a fit, experimentalists obtain $m(D')=2539(8)$ MeV and $\Gamma(D')=130(18)$ MeV.
A first question raised about the fairness of this interpretation because, on the theoretical side, quark models predicted roughly the same $D'$ mass (2.58 GeV) but a quite smaller width (70 MeV) \cite{quarkmodels}. However there is a well known caveat here: excited states properties are very sensitive to the position of the wave functions nodes, that actually depend strongly on the quark model.
Examinating the semileptonic decay $B \to D' l \nu$, assuming it is quite large \cite{bernlochner} and using the fact that $\Gamma (D' \to D_{1/2} \pi) \gg \Gamma(D' \to D_{3/2} \pi)$\footnote{Spectroscopy notations: $D_{1/2} \equiv \{D^*_0,\, D^*_1\}$, $D_{3/2}\equiv\{D_1,\, D^*_2\}$.}, one arrives at the conclusion that an excess of $B \to (D_{1/2}\pi) l \nu$ events could be observed with respect to their $B \to (D_{3/2} \pi) l \nu$ counterparts. One may then wonder whether such a potentially large $B \to D' l \nu$ width could explain the "1/2 vs. 3/2" puzzle: $[\Gamma (B \to D_{1/2} l \nu) \simeq \Gamma (B \to D_{3/2} l \nu)]^{\rm exp}$ while 
$[\Gamma (B \to D_{1/2} l \nu) \ll \Gamma (B \to D_{3/2} l \nu)]^{\rm theory}$ \cite{puzzle}.
Finally there is still a $\sim 3 \sigma$ discrepancy between the exclusive determination of the CKM matrix element $V_{cb}$ and its inclusive determination, mainly due to a very small error on both sides. $|V_{cb}|^{\rm excl}$ is extracted from $B \to D^{(*)} l \nu$ decays and needs, at a normalization point, the theoretical computation of the form factors associated to $B \to D^{(*)}$ transitions. An analysis performed in the OPE formalism argued that a large $B \to D'$ form factor is going together with a small suppression of the $B \to D^{(*)}$ counterpart \cite{gambinouraltsev}: one may ask whether it could involve a reduction of the discrepancy between $|V_{cb}|^{\rm excl}$ and $|V_{cb}|^{\rm incl}$ \cite{vckmfit}.

\section{Non leptonic $B \to D'$ decay}

We have proposed in \cite{BecirevicMP} to check the hypothesis of a large branching ratio ${\cal B}(B \to D' l \nu)$ by studying non leptonic decays. First, considering the Class I process $\bar{B}^0 \to D'^+ \pi^-$, one has in the factorisation approximation
\begin{equation}
\frac{{\cal B}(\bar{B}^0 \to D'^+ \pi^-)}
{{\cal B}(\bar{B}^0 \to D^+ \pi^-)}=
\left(\frac{m^2_B - m^2_{D'}}{m^2_B - m^2_{D}}\right)^2 
\left[\frac{\lambda(m_B,m_{D'},m_\pi)}{\lambda(m_B,m_{D},m_\pi)}\right]^{1/2}
\left|\frac{f^{B \to D'}_+(0)}{f^{B \to D}_+(0)}\right|^2,
\end{equation}
\begin{equation}
\nonumber
\lambda(x,y,z)=[x^2 - (y + z)^2][x^2 - (y - z)^2], \quad
f^{B \to D(')}_+(m^2_\pi) \sim f^{B \to D(')}_+(0).
\end{equation}
Using $V_{cb}f^{B \to D}_+(0)=0.02642(8)$ from Babar \cite{vcbfpbabar} and 
$\left|V_{cb}\right|^{\rm incl} = 0.0411(16)$, we deduce $f^{B \to D}_+(0) = 0.64(2)$. Then, 
with $m_{D'}=2.54$ GeV, we obtain $\frac{{\cal B}(\bar{B}^0 \to D'^+ \pi^-)}
{{\cal B}(\bar{B}^0 \to D^+ \pi^-)}= (1.65 \pm 0.13) \times
\left|f^{B \to D'}_+(0)\right|^2$. With ${\cal B}(\bar{B}^0 \to D^+ \pi^-)=0.268(13)\%$, we have finally
\begin{equation}
{\cal B}(\bar{B}^0 \to D'^+ \pi^-) = \left|f^{B \to D'}_+(0)\right|^2 \times
(4.7 \pm 0.4) \times 10^{-3}.
\end{equation}
Letting vary the $f_+^{B \to D'}(0)$ form factor in a quite large range [0.1, 0.4], according to the existing theoretical estimates \cite{bernlochner}, \cite{fBDpanel}, we conclude that ${\cal B}(\bar{B}^0 \to D'^+ \pi^-)^{\rm th} \sim 10^{-4}$: \emph{it can be measured with the B factories samples and at LHCb.}

Second, investigating the Class III process $B^- \to D'^0 \pi^-$, we write the factorised amplitude in the following way:
\begin{equation}
\nonumber
A^{III}_{\rm fact} = -i \frac{G_F}{\sqrt{2}}
V_{cb}V^*_{ud} \left[a_1f_\pi
[m^2_B - m^2_{D'}]f^{B \to D'}(m^2_\pi)
+a_2f_{D'}[m^2_B - m^2_\pi]f^{B \to \pi}(m^2_{D'})\right].
\end{equation}
Normalising the corresponding branching ratio by the Class I counterpart, we get
\begin{equation}\nonumber
\frac{{\cal B}( B^-\to D^{\prime 0}\pi^-)}{{\cal B}(\bar{B}^0\to  D^{\prime +}
\pi^-)} =\frac{\tau_{B^-}}{\tau_{\bar B^0}}\left[ 1 + 
{a_2\over a_1}\times 
{m_B^2-m_\pi^2\over m_B^2-m_{D^\prime}^2} 
 \times  
{f_0^{B\to \pi}(m_{D^\prime}^2)\over f_+^{B \to D'}(0)}  
{f_{D^\prime}\over f_D} \ {f_D\over f_\pi}\right]^2.
\end{equation}
The ratio of Wilson coefficients $a_2/a_1$ is determined from $\frac{{\cal B}( B^-\to D^0\pi^-)}
{{\cal B}(\bar{B}^0\to  D^+ \pi^-)}$, known experimentally \cite{vckmfit}, and it remains to compute
on the lattice the ratios of decay constants $\frac{f_{D'}}{f_D}$ and $\frac{f_D}{f_\pi}$.

\section{Lattice calculation}

\begin{figure}
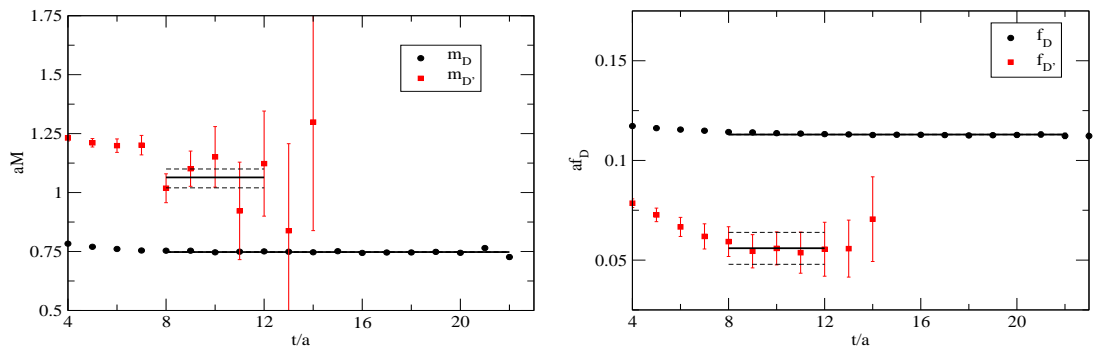

\begin{center}
\begin{tabular}{cc}
\includegraphics*[width=7cm, height=4.5cm]{plots/figmDmDp.eps}
&
\includegraphics*[width=7cm, height=4.5cm]{plots/figfDfDp.eps}\\
\end{tabular}
\end{center}
\caption{\label{figplatmassfD} Plateaus of $m_D$ and $m_{D'}$ (left panel), $f_D$ and $f_{D'}$ (right panel) for the ETMC ensemble $\beta=3.9, \mu_{\rm sea}=0.0064$.}
\end{figure}

\begin{figure}
\begin{center}
\includegraphics*[width=7cm, height=4.5cm]{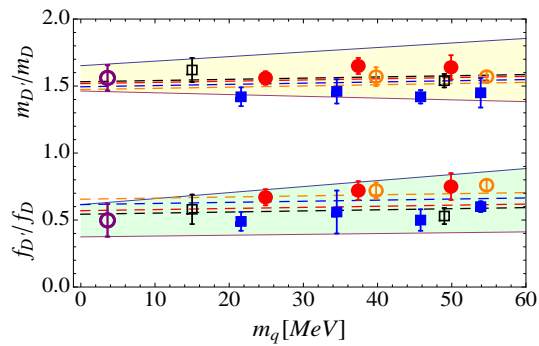}
\end{center}
\caption{\label{figfitfDpmDp} Chiral and continuum limit extrapolations of $m_{D'}/m_D$ and $f_{D'}/f_D$.}
\end{figure}

Our analysis is based on simulations built by the ETM Collaboration \cite{etmc} with ${\rm N_f}=2$ fermions regularised with the Twisted-mass QCD action tuned at maximal twist. At a fixed lattice spacing we crosscheck our results on a simulation performed by the QCDSF collaboration with ${\rm N_f=2}$ Wilson-Clover fermions \cite{qcdsf} and a third one realised in the quenched approximation. We collect the simulations parameters in Tables \ref{tab:01} and \ref{tab:01bis}.
\begin{table*}[h!!]
\begin{center}
\begin{tabular}{|c|cccccc|}   
\hline
$ \beta$& 3.8 &  3.9  &  3.9 & 4.05 & 4.2  & 4.2    \\ 
$ L^3 \times T $&  $24^3 \times 48$ & $24^3 \times 48$  & $32^3 \times 64$ & $32^3 \times 64$& $32^3 \times 64$  & $48^3 \times 96$  \\ 
$ \#\ {\rm meas.}$& 240 &  240 & 150 & 150 & 150 & 100  \\ \hline 
$\mu_{\rm sea 1}$& 0.0080 & 0.0040 & 0.0030 & 0.0030 & 0.0065 &  0.0020   \\ 
$\mu_{\rm sea 2}$& 0.0110 & 0.0064 & 0.0040 & 0.0060 &   &     \\ 
$\mu_{\rm sea 3}$&  &   &  & 0.0080 &   &     \\  \hline
$a \ {\rm [fm]}$&   0.098(3) & 0.085(3) & 0.085(3) & 0.067(2) & 0.054(1) & 0.054(1)      \\ 
$\mu_{s}$& 0.0194(7)  &0.0177(6)  &0.0177(6)   & 0.0154(5) & 0.0129(5) & 0.0129(5)  \\ 
$\mu_{c}$& 0.2331(82)  &0.2150(75)  &0.2150(75)   & 0.1849(65) & 0.1566(55) & 0.1566(55)  \\ 
\hline
\end{tabular}
\end{center}
{\caption{\footnotesize  \label{tab:01} Lattice ensembles used in this work with the indicated number of gauge field configurations. Lattice spacing is set by using the Sommer parameter $r_0/a$, with $r_0= 0.440(12)$~fm fixed by matching $f_\pi$ obtained on the lattice with its physical value  (cf. ref.~\cite{BlossierCR}). Quark mass parameters $\mu$ are given in lattice units.}}
\vspace{0.5cm}

\begin{center}
\begin{tabular}{|ccccccc|}   
\hline
${\rm N_f}$ & $ \beta$ ($c_{SW}$) & $ L^3 \times T $ & $ \#\ {\rm meas.}$ &  $\kappa_{\rm sea}$ & $\kappa_{\rm s}$ & $\kappa_{\rm c}$   \\  \hline
0 & 6.2 (1.614) & $24^3 \times 48$ & 200 &  -- & 0.1348 & 0.125 \\ 
2 & 5.4 (1.823) & $24^3 \times 48$ & 160 & 0.13625 & 0.1359 & 0.126 \\ 
\hline
\end{tabular}
\end{center}
{\caption{\footnotesize  \label{tab:01bis} Lattice set-up for the results obtained by using the Wilson gauge and the Wilson-Clover quark action. $\kappa_{\rm sea}$, $\kappa_s$ and $\kappa_c$ stand for the value of the hopping parameter of the sea, strange and the charm quark respectively.}}
\end{table*}
A first way to extract $m_{D'_q}$ and $f_{D'_q}$ is to consider the two-point correlator
\begin{eqnarray}
\nonumber
C_{D_qD_q}(t)&=& \langle {\displaystyle \sum_{\vec x} }  P_{D_q}(\vec x; t)  
P_{D_q}^\dagger (0; 0) \rangle\\  
\nonumber
&\xrightarrow[]{\displaystyle{ t\gg 0}}&
 \;  \left| {\cal Z}_{D_q} \right|^2 
 \frac{\cosh[  m_{D_q} (T/2-t)]}{m_{D_q} } 
 e^{- m_{D_q} T/2}.
 \end{eqnarray}
We define a modified two-point correlation function by subtracting the ground state contribution:
\begin{equation}
\nonumber
C_{D_{q}D_{q}}^\prime (t) = C_{D_{q}D_{q}}(t) - 
\left| {\cal Z}_{D_q} \right|^2 \frac{\cosh[  m_{D_q} (T/2-t)]}{ m_{D_q} } 
e^{- m_{D_q} T/2};
\end{equation}
we extract the effective mass $m_{D'_q}$ from the ratio $\frac{C_{D_qD_q}^{\prime}(t)}
{C_{D_qD_q}^{\prime}(t+1)}=\frac{\cosh\left[ m_{D_q^\prime}^{\rm eff}(t) 
\left( {\displaystyle{T\over 2}} - t\right)\right]}{ 
\cosh\left[m_{D_q^\prime}^{\rm eff}(t) \left( 
{\displaystyle{T\over 2}} - t -1\right)\right]}$ and the decay constant $f_{D'_q}$ from a fit of $C_{D_qD_q}^{\prime}$.\\
An alternative approach consists in using a basis of interpolating fields $P_{D_q\,i} \equiv \bar{\psi}_{c\,i}\gamma^5 \psi_{q\,i}$ by smearing with a "Gaussian" wave function the local fields $\psi_{c(q)}$: $\psi_{c(q)\,i}=\left(\frac{1 + \kappa_G H}{1 + 6\kappa_G}\right)^{n_i}\psi_{c(q)}$,\\ 
with $H_{i,j}=\sum_{\mu=1}^3\left(U^{n_a}_{i;\mu}\delta_{i+\mu,j}+U^{n_a\dagger}_{i-\mu;\mu}\delta_{i-\mu,j}\right)$;
$U^{n_a}_{i,\mu}$ is a $n_a$ times APE smeared link.
We solve the generalized eigenvalue problem (GEVP):
\begin{equation}
\nonumber
C_{D_{q}D_{q}\;ij}(t) v^{(n)}_j(t,t_0)=\lambda^{(n)}(t,t_0)
C_{D_{q}D_{q}\;ij}(t_0) v^{(n)}_j(t,t_0).
\end{equation}
The projected creation operator that has the largest coupling to the $n^{th}$ excited state is defined by\begin{equation}
\widetilde{P}^{(n)}_{D_{q}}(t,t_{0})=\sum_{i} v^{(n)}_{i}(t,t_{0})P_{D_{q}\,i} \quad
\langle  D_q^{(m)} 
\vert \widetilde{P}^{(n)\dag}_{D_{q}}\vert 0\rangle =A_n \delta_{mn}.
\end{equation}
Effective masses and decay constants read
\begin{equation}
\nonumber
m_{D_q^{(n)}}^{\rm eff}(t)={\rm arccosh}\left[ 
{\lambda^{(n)}(t+1,t_0) + \lambda^{(n)}(t-1,t_0)\over 2 
\lambda^{(n)}(t,t_0)}\right], \quad
\langle  D_q^{(n)} \vert P_{D_{q}\,L}^\dag \vert 0\rangle =
{\sqrt{A_n}~~\sum_i C_{D_{q}D_{q}\;Li}(t)v^{(n)}_i(t,t_0)\over 
\displaystyle  \sum_{ij}
v^{(n)}_i(t,t_0)C_{D_{q}D_{q}\;ij}(t)
v^{(n)}_j(t,t_0)},
\end{equation}
where the label "L" refers to a local interpolating field. In our simulation we have chosen $\kappa_G=4.0$ and we have considered a basis of 4 operators with $n_i=\{0,2,10,30\}$.
We show in Figure \ref{figplatmassfD} plateaus of $D$ and $D'$ masses and decay constants for one of the ETMC ensembles.
We perform a combined chiral and continuum extrapolation of $m_{D'_q}/m_{D_q}$ and 
$f_{D'_q}/f_{D_q}$ with the formula
\begin{equation}\label{eq:fDpmDp}
{\cal F}^{\rm latt.} = A_{\cal F} \left[ 
1 + B_{{\cal F}} m_q + C_{{\cal F}} \left( {a\over a_{\beta=3.9}}\right)^2\right].
\end{equation}
The fits quality is illustrated in Figure \ref{figfitfDpmDp}.
Our results read:
\begin{equation}
{m_{D_s^\prime}\over m_{D_s}} = 1.53(7), \quad {f_{D_s^\prime}\over f_{D_s}} = 0.59(11),\quad
{m_{D^\prime}\over m_{D}} = 1.55(9), \quad {f_{D^\prime}\over f_{D}} =0.57(16).
\end{equation}
There is a $\sim 2\sigma$ discrepancy with the experimental estimate ${m_{D^\prime}\over m_{D}} = 1.36$.
To check whether it could arise because Twisted-mass QCD breaks parity at finite lattice spacing, inducing a mixing between radial excitations and states of opposite parity that would not be properly taken into account in our work, we perform a computation with Wilson-Clover fermions at a lattice spacing corresponding to $a_{\beta=4.05}$ (cf. Tables \ref{tab:01} and \ref{tab:01bis}). Another source of systematics has a physical origin: with ${\rm N_f}=2$ dymamical light quarks, decay channels can open up and transitions $D' \to D^* \pi$, $D' \to D^*_0 \pi$, 
$D'_s \to D^* K$ and $D'_s \to D^*_0 K$ are kinematically allowed in large volumes, making the analysis in principle very tricky. As it is not the case in the quenched approximation, we checked our findings in that framework as well. We observe in Table \ref{tabcomp} a qualitative good agreement between our estimates of $m_{D'_s}/m_{D_s}$ and $f_{D'_s}/f_{D_s}$ at finite lattice spacing.
\begin{table}[t]
\begin{center}
\begin{tabular}{|c|c|c|}   
\cline{2-3}
\multicolumn{1}{l|}{}&$m_{D'_s}/m_{D_s}$&$f_{D'_s}/f_{D_s}$\\
\hline
tmQCD (${\rm N_f}=2$)&1.55(6)&0.69(5)\\
Wilson-Clover (${\rm N_f}=2$)&1.48(7)&0.77(9)\\
Wilson-Clover (${\rm N_f}=0$)&1.41(9)&0.67(12)\\
\hline
\end{tabular}
\end{center}
\caption{\label{tabcomp} Comparison of results obtained at $a \sim 0.065$ fm with different quark regularisations and numbers of dynamical flavours.}
\end{table}
Moreover we extrapolate to the physical point $f_{D_{s}}/m_{D_{s}}$ with the formula (\ref{eq:fDpmDp}) and $\sqrt{m_{D_s}/m_D} [f_{D_s}/f_D]/[f_K/f_\pi]$ using Heavy Meson Chiral Perturbation Theory at LO ($X=0$) and NLO ($X=1$) \cite{HMChPT}, deducing $f_{D}/f_{\pi}$:
\begin{equation}
\sqrt{m_{D_s}/m_D} [f_{D_s}/f_D]/[f_K/f_\pi]=A \left[  1 + X  {9 g^2-4\over 4(4\pi f)^2} m_\pi^2 \log(m_\pi^2)  
+ B m_\pi^2 + C \left( {a\over a_{\beta=3.9}}\right)^2\right],
\end{equation}
with $g=0.53(3)(3)$ \cite{our-g},
\begin{equation}
f_{D_s}/m_{D_s}=0.1281(11),\quad f_{D_s}/f_D=1.23(1)(1), \quad f_D/f_\pi=1.56(3)(2),
\end{equation}
where the first error is statistical and the second error corresponds to the difference between LO and NLO chiral fits of $\sqrt{m_{D_s}/m_D}[f_{D_s}/f_D]/[f_K/f_\pi]$. Finally we obtain $f_{D_{s}}=252(3)$~MeV, in excellent agreement with a very recent measurement at Belle: $f^{\rm exp}_{D_{s}}=255 \pm 4.2 \pm 5.1$~MeV \cite{BellefDs}.

\section{Back to phenomenology and conclusion}

We have now everything we need to answer our question at the beginning. With $a_{2}/a_{1}=0.368$, $\tau_{\bar B^0}/\tau_{B^-}=1.079(7)$, $f_+^{B\to D}(0)= 0.64(2)$ and $f_0^{B\to \pi}(m_D^2)= 0.29(4)$ \cite{blazenka}, we find
\begin{equation}\nonumber
{{\cal B}( B^-\to D^{\prime 0}\pi^-)\over {\cal B}(\bar{B}^0\to  D^{\prime +}
\pi^-)} ={\displaystyle{\tau_{B^-}\over \tau_{\bar B^0}} \left[ 1 
+ \displaystyle{0.14(4)\over f_+^{B \to D'}(0)} \right]^2}, \quad
{{\cal B}(\bar{B}^0\to D^{\prime +}\pi^-)
\over {\cal B}(\bar{B}^0\to D^{+}\pi^-)} =
(1.24\pm 0.21)\times |f_+^{B \to D'}(0)|^2.
\end{equation}
Using the experimental value $\frac{m_{D^\prime}}{m_{D}}=1.36$, we find $\frac{{\cal B}(\bar{B}^0 \to D'^+ \pi^-)}
{{\cal B}(\bar{B}^0 \to D^+ \pi^-)}=(1.65 \pm 0.13) \times
\left| f^{B \to D'}_+(0)\right |^2$: in other words, the dependence on $m_{D'}$ of that ratio is small. 
Setting $f^{B \to D'}_+=0.4$ as found by Ebert \emph{et al} \cite{fBDpanel}, 
$m_{D'}/m_D=1.36$ and the branching ratio 
${\cal B}(\bar{B}^0\to  D_2^{\ast +}\pi^-)$
measured at $B$ factories \cite{B0D2pi}, \cite{BmD2pi}, we obtain
\begin{equation}
\frac{{\cal B}( \bar{B}^0\to  D^{\prime +}\pi^-)}{{\cal B}(\bar{B}^0\to  D_2^{\ast +}\pi^-)}
=1.6(3), \quad
\frac{{\cal B}( B^-\to D^{\prime 0}\pi^-)}{{\cal B}(B^- \to  D_2^{\ast 0}\pi^-)} =1.4(3).
\end{equation}
In conclusion, if $f^{B \to D'}$ is large, as claimed by many authors, the measurement of
${\cal B}(B \to D' \pi)$ should be as feasible in present experiments as ${\cal B}(B \to D^{*}_{2} \pi)$ was
at $B$ factories. Thus, it reveals beneficial to study $B \to D'$ non leptonic decays to address the composition
of final states in $B \to D^{**} l \nu$ semileptonic decays. A natural extension of our work is to consider the process $B \to D^*_0 \pi$ and compute on the lattice $f_{D^*_0}$.


\begin{thebibliography}{99}
\bibitem{babard}
  P.~del Amo Sanchez {\it et al.}  [BABAR Collaboration],
  Phys.\ Rev.\ D {\bf 82} (2010) 111101.

\bibitem{quarkmodels}
F.~E.~Close and E.~S.~Swanson,
  Phys.\ Rev.\ D {\bf 72} (2005) 094004;
  Z.~-F.~Sun, J.~-S.~Yu, X.~Liu and T.~Matsuki,
  Phys.\ Rev.\ D {\bf 82} (2010) 111501.

\bibitem{bernlochner}
  F.~U.~Bernlochner, Z.~Ligeti and S.~Turczyk,
  Phys.\ Rev.\ D {\bf 85} (2012) 094033.

\bibitem{puzzle}
 A.~Le Yaouanc \emph{et al},
  Phys.\ Rev.\ D {\bf 56} (1997) 5668;
  A.~K.~Leibovich \emph{et al},
  Phys.\ Rev.\ D {\bf 57} (1998) 308;
D.~Becirevic \emph{et al},
  Phys.\ Rev.\ D {\bf 87}, no. 5, 054007 (2013);
  I.~I.~Bigi \emph{et al},
  Eur.\ Phys.\ J.\ C {\bf 52}, 975 (2007).

\bibitem{gambinouraltsev}
  P.~Gambino, T.~Mannel and N.~Uraltsev,
  JHEP {\bf 1210} (2012) 169.
  
\bibitem{vckmfit} 
J.~Beringer {\it et al.}  [Particle Data Group Collaboration],
Phys.\ Rev.\ D {\bf 86}, 010001 (2012).

\bibitem{BecirevicMP}
  D.~Becirevic \emph{et al},
Nucl.\ Phys.\ B {\bf 872}, 313 (2013).
\bibitem{vcbfpbabar}
  B.~Aubert {\it et al.}  [BABAR Collaboration],
  Phys.\ Rev.\ Lett.\  {\bf 104} (2010) 011802.

\bibitem{fBDpanel}
  J.~Hein {\it et al.}  [UKQCD Collaboration],
  Nucl.\ Phys.\ Proc.\ Suppl.\  {\bf 83} (2000) 298;
  D.~Ebert, R.~N.~Faustov and V.~O.~Galkin,
  Phys.\ Rev.\ D {\bf 62} (2000) 014032;
R.~N.~Faustov and V.~O.~Galkin,
  Phys.\ Rev.\ D {\bf 87}, 034033 (2013);
 Z.~-H.~Wang \emph{et al},
  J.\ Phys.\ G {\bf 39} (2012) 085006.

\bibitem{etmc}
P.~Boucaud {\it et al.}  [ETM Collaboration],
  Phys.\ Lett.\ B {\bf 650} (2007) 304; 
  Comput.\ Phys.\ Commun.\  {\bf 179} (2008) 695.

\bibitem{qcdsf}
A.~Ali Khan {\it et al.}  [QCDSF Collaboration],
  Phys.\ Lett.\ B {\bf 564} (2003) 235;
M.~Gockeler {\it et al.}  [QCDSF Collaboration],
  Phys.\ Rev.\ D {\bf 82} (2010) 114511
   [Erratum-ibid.\ D {\bf 86} (2012) 099903].

\bibitem{BlossierCR}
  B.~Blossier {\it et al.}  [ETM Collaboration],
Phys.\ Rev.\ D {\bf 82}, 114513 (2010).



\bibitem{HMChPT}
  R.~Casalbuoni \emph{et al},
  Phys.\ Rept.\  {\bf 281} (1997) 145;
  G.~Burdman and J.~F.~Donoghue,
  Phys.\ Lett.\  B {\bf 280} (1992) 287.


\bibitem{our-g}
  D.~Becirevic and F.~Sanfilippo,
  Phys.\ Lett.\ B {\bf 721}, 94 (2013).

\bibitem{BellefDs}
  A.~Zupanc {\it et al.}  [Belle Collaboration],
JHEP {\bf 1309}, 139 (2013).

\bibitem{blazenka}
  G.~Duplancic \emph{et al},
  JHEP {\bf 0804} (2008) 014;
P.~Ball and R.~Zwicky,
  Phys.\ Rev.\ D {\bf 71} (2005) 014015.

\bibitem{B0D2pi}
P.~del Amo Sanchez {\it et al.}  [BABAR Collaboration],
  PoS ICHEP {\bf 2010} (2010) 250;
  Phys.\ Rev.\ D {\bf 76} (2007) 012006.

\bibitem{BmD2pi}
B.~Aubert {\it et al.}  [BABAR Collaboration],
  Phys.\ Rev.\ D {\bf 79} (2009) 112004;
K.~Abe {\it et al.}  [Belle Collaboration],
  Phys.\ Rev.\ D {\bf 69} (2004) 112002.

\end{thebibliography}
\end{document}